\documentclass[prb,twocolumn,showpacs,preprintnumbers,amsmath,amssymb]{revtex4}

\usepackage{graphicx}
\usepackage{dcolumn}
\usepackage{bm}
\usepackage{tabularx}
\usepackage{amsmath}

\begin{document}


\title{Thermal Conductivity across the Phase Diagram of Cuprates: 

Low-Energy Quasiparticles and Doping Dependence of the Superconducting Gap}

\author{Mike Sutherland, D.G. Hawthorn,  R.W. Hill, F. Ronning, S. 
Wakimoto, H. Zhang, C. Proust$^{\star}$,  
Etienne Boaknin, C. Lupien$^{\dagger}$ and Louis Taillefer}

\affiliation{} 

\affiliation{Canadian Institute for Advanced Research, Department of Physics, University of
Toronto, 60 St. George Street, Toronto, Ontario, Canada M5S 1A7}

\author{Ruixing Liang, D.A. Bonn, and W.N. Hardy}
\affiliation{Department of Physics, University of British Columbia, 
6224 Agricultural Road, Vancouver, B.C. Canada V6T 1Z1}

\author{Robert Gagnon}
\affiliation{Department of Physics, McGill University, 3600 University Street, Montr\'eal, Queb\'ec, Canada H3A 
2T8}

\author{N.E. Hussey}
\affiliation{Department of Physics, University of Bristol, Tyndall Avenue, Bristol, BS8 1TL, United Kingdom}

\author{T. Kimura, M. Nohara, H. Takagi}
\affiliation{Department of Advanced Materials Science, Graduate School of Frontier Sciences, University of 
Tokyo, Hongo 7-3-1, Bunkyo-ku, Tokyo 113-8656, Japan}

\date{\today}

\begin{abstract}

Heat transport in the cuprate superconductors YBa$_2$Cu$_3$O$_{y}$ and La$_{2-x}$Sr$_x$CuO$_4$ was measured 
at low temperatures as a function of doping. A residual linear term $\kappa_{0}/T$ is observed throughout the superconducting
region and it decreases steadily as the Mott insulator is approached from the overdoped regime.
The low-energy quasiparticle gap extracted from $\kappa_{0}/T$ is seen to scale
closely with the pseudogap. The ubiquitous presence of nodes and the tracking of the pseudogap shows that
the overall gap remains of the pure $d$-wave form throughout the phase diagram, which excludes the 
possibility of a complex component ($ix$) appearing at a putative quantum phase transition and argues 
against a non-superconducting origin to the pseudogap. A comparison with superfluid density measurements 
reveals that the quasiparticle effective charge is weakly dependent on doping and close to unity.

\end{abstract}

\pacs{74.25.Fy, 74.72.Bk, 74.72.Dn}

\maketitle

\section{\label{sec:intro}Introduction}

In a $d$-wave superconductor, the presence of nodes in the gap structure imposed by symmetry leads to
quasiparticle excitations down to zero energy in the presence of even small amounts of disorder.
\cite{Hirschfeld,Schmitt-Rink}
These excitations are delocalized and carry both charge and heat.
The most striking property of this residual normal fluid is its universal conduction, \cite{Lee} 
whereby quasiparticle transport is independent of impurity concentration. In the case of heat transport,
it turns out to be a direct measure of the low-energy quasiparticle spectrum. \cite{Durst} The universal 
character of heat transport
was confirmed experimentally for the cuprates YBa$_2$Cu$_3$O$_{y}$ (YBCO) \cite{Taillefer} and
Bi$_2$Sr$_2$CaCu$_2$O$_8$ (Bi-2212). \cite{Nakamae}  Moreover, the residual heat conduction measured at 
optimal doping \cite{Chiao} or above, in the overdoped regime, \cite{Proust} 
is in good quantitative agreement with $d$-wave BCS theory and the 
quasiparticle spectrum either measured by angle-resolved photoemission spectroscopy (ARPES)
or expected from estimates based on the value of $T_c$ (see for instance Ref.~7).

In this paper, we use the well-established and robust connection between low-temperature heat transport and the energy
spectrum of a $d$-wave superconductor to probe the evolution of low-energy quasiparticles and the superconducting gap as a 
function of doping in the cuprates. In going from the overdoped to the underdoped 
regime, we find that the residual linear term $\kappa_{0}/T$ is finite everywhere and decreases monotonically. As a result the 
low-energy gap grows steadily, in contrast to the superconducting $T_c$ which first rises and then decreases. The low-energy gap in fact 
follows closely the normal-state pseudogap, measured mostly at higher energies and 
temperatures.  \cite{statt} 
The ubiquitous presence of nodes and the tracking of the pseudogap shows that 
the gap remains of the pure $d$-wave form throughout the phase diagram. This excludes the possibility of a 
complex component ($ix$) appearing at a putative quantum phase transition and argues against
a non-superconducting origin to the pseudogap.

\section{\label{Samples}Samples}

We performed our study on two cuprate materials: the double-plane orthorhombic material YBa$_2$Cu$_3$O$_{y}$ (YBCO) 
doped with oxygen in CuO chains, and
 the single-plane material La$_{2-x}$Sr$_x$CuO$_4$ doped with Sr 
atoms (LSCO). 
The four samples of YBCO used in the study are detwinned, 
flux-grown single crystals in the shape of platelets
with typical dimensions  $1.0\times0.5$ mm and 25 $\mu$m thick.  Two of them, respectively with
$y = 6.99$ and $y = 6.54$, were grown in a BaZrO$_3$ (BZO) crucible 
,\cite{Liang} which results in crystals with
very high chemical purity (99.99 -- 99.995\%) and a high degree of
crystalline perfection as compared with crystals grown in Y$_2$O$_3$-stabilised ZrO$_2$ (YSZ) crucibles. 
The sample with $y = 6.99$ was detwinned at 
250$^\circ$C under uniaxial stress, and then annealed at 350$^\circ$C for 50 days, resulting 
in CuO chains with less than 0.2\% oxygen vacancies, and hence very close to the stoichiometric composition
at $y = 7.00$. \cite{Liang}  This level of oxygen doping is slightly above that for maximal $T_c$ (93.6 
K),  resulting in $T_c = 89$ K.  The sample with $y = 6.54$ was similarly detwinned and then annealed at 
84$^\circ$C for 2 days followed by 
60$^\circ$C for 5 days.  This 
resulted in a highly-ordered ortho-II arrangement of oxygen atoms in CuO chains, with alternating full and empty chains.
The other two YBCO samples, respectively with
$y = 6.95$ and $y = 6.6$, were grown in a YSZ crucible, and are characterized by an impurity concentration typically one 
order of magnitude higher. The oxygen vacancies in the CuO chains are not ordered for these crystals.  The 
$y = 6.6$ sample was quenched into an ice water bath after annealing, resulting in a higher level of disorder 
among the oxygen vacancies and thus a lower $T_c$ as compared to non-quenched samples with similar oxygen 
content.


The La$_{2-x}$Sr$_x$CuO$_4$ (LSCO) samples were all grown in an image furnace using the
travelling-solvent floating-zone technique and have Sr dopings of $x$~= 0.06 (samples A and B), 0.07, 0.09, 0.17 and 0.20.  
In addition, a non-superconducting LSCO sample with $x=0.05$ was also measured.  With the exception of $x=0.06$ B, 
all samples were measured as grown.  This may result in off-stoichiometric oxygen content in the samples.  The 
$x=0.06$ B sample was annealed in flowing argon overnight 
at 800$^\circ$C in an attempt to fix the oxygen content.  
The argon annealing, however, had little effect on our results as both $x=0.06$ samples gave the 
same electronic contribution to the thermal conductivity.

In LSCO, the hole concentration per Cu in the $\rm{CuO_2}$ planes, $p$, is taken to be the Sr concentration, $x$.  
In YBCO, however, the relation between hole concentration and oxygen doping $y$ is a complicated function. 
As a result, for YBCO $p$ is determined from transition temperatures using the empirical formula 
\cite{Presland}

\begin{equation}
\frac{T_c}{T_c^{max}}=1-82.6(p - 0.16)^2,
\label{eq:Tallon}
\end{equation}
which is a good approximation for many cuprate systems. \cite{Tallon}  Here we use $T_c^{max}$ = 93.6 
K as the transition temperature of 
optimally-doped YBCO.

\begin{figure}
\vspace{8pt}
\centering
\resizebox{\columnwidth}{!}{\includegraphics{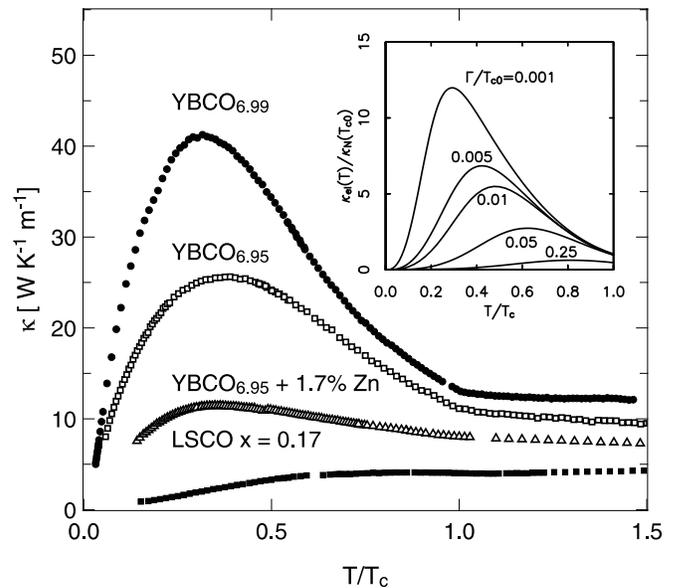}}
\caption{\label{fig:dipperkappa}Thermal conductivity of $\rm
YBa_2Cu_3O_{y}$ and La$_{2-x}$Sr$_x$CuO$_4$ vs temperature normalized at
$T_c$.
{\sl Inset}: theoretical calculation of the effect of impurity scattering
on the electronic thermal conductivity of cuprates (from Ref.~17).}
\end{figure}

The transition temperature was
determined from resistivity measurements and defined as the temperature 
at which the resistivity has fallen to zero.  Note that this definition of $T_c$ leads to values slightly 
lower than those determined by taking the 
midpoint of the resistive transition. We find however that values of $T_c$ determined this way 
correlate well with those measured by magnetic 
susceptibility and thermal conductivity.   
For the YBCO samples, the transition temperatures are 62, 44, 93.5 and 89 K, respectively, 
for oxygen doping $y = 6.54, 6.6, 6.95$ and 6.99.  We note that the $T_c$ 
for the ortho-II ordered $y=6.54$ appears anomalously high. 
This enhanced $T_c$ (consistent with 
the $T_c$ determined by magnetization measurements in similar samples \cite{liang2}) is thought to 
be due to oxygen 
coordination effects, where improved oxygen ordering leads to a greater number of holes doped into the CuO$_2$ planes 
as compared to a non-ordered crystal with the same nominal oxygen doping. \cite{Veal}
For the LSCO samples, the transition temperatures are 5.5, 8.5, 19, 16, 
34 and 33.5 K, respectively, 
for Sr doping $x = 0.06$ (sample A, not annealed), 0.06 (sample B, 
annealed), 0.07, 0.09, 0.17 and 0.20.  The $T_{c}$ for the 
$x=0.09$ sample is anomalously low, possibly as a result of Sr 
inhomogeneity or oxygen non-stoichiometry within the crystal.  Although  
different criteria may be used for determining the value of hole doping level in both the LSCO 
and YBCO systems, we note that small errors in our estimation of hole concentration do not 
noticeably affect the trends observed in our thermal conductivity data. 

The low temperature thermal conductivity measurements were
made in a dilution refrigerator down to 40 mK, using the standard 4-wire steady-state method with 
two $\rm{RuO_2}$ chip thermometers, calibrated in-situ
against a reference Ge thermometer. Currents were applied along the $a$-axis in order to probe in-plane transport and
to avoid contributions from the CuO chains in YBCO. Thermal and electrical contact to the samples was made 
using Ag-wire and diffused Epotek H20E Ag-epoxy pads.

\section{\label{Disorder}Levels of disorder}

It is instructive to estimate the relative amounts of disorder in our 
samples.  For the YBCO samples, the inclusion of impurities during growth is greatly reduced by using 
BZO over YSZ crucibles. Microwave spectroscopy measurements of thermally-excited
quasiparticles in the elastic scattering limit reveal that the scattering rate in the superconducting 
state is some 12 times greater for optimally-doped 
YBCO grown in YSZ crucibles \cite{Hardy}  compared to the slightly overdoped $y=6.99$ samples grown in 
BZO crucibles. \cite{Hosseini} 
Measurements of thermal conductivity $\kappa$ at high temperature, shown in Fig.~1, lead to a similar order-of-magnitude
difference.  In the theory of Hirschfeld and Putikka \cite{Hirschfeld2}, 
the peak observed in the thermal conductivity is due to an 
increase in the quasiparticle mean free path when the sample is cooled below $T_c$. The magnitude of $\kappa$ continues to increase with 
cooling until it becomes limited by quasiparticle scattering from impurities and dislocations.  Thus, the ratio of
peak height to normal-state value in $\kappa(T)$ directly reflects the amount of disorder present in the crystal.  
The inset in Fig.~1 shows theoretical curves that demonstrate this effect, 
\cite{Hirschfeld2} where the 
electronic contribution to thermal conductivity normalized by the value of $\kappa$ at $T_{c}$ is plotted as a function of 
$T/T_{c}$.  

 Samples with a large ratio of impurity scattering rate (given by $\Gamma$) to $T_{c0}$ are seen to have a peak height that is 
suppressed.  An order of magnitude increase in the intrinsic level of disorder within the crystal results roughly in a 
factor of two decrease in the peak height.  The 
data presented for our crystals in Fig.~1 reflects that effect.  The deliberate addition of impurities such as
Zn in YSZ-grown optimally-doped YBCO samples leads to a large suppression in 
peak height. Data for a sample with 1.7\% Zn impurities (determined from $T_c$ suppression) is 
shown in Fig.~1, and it is seen that the addition of this level of impurities causes the 
peak to nearly vanish. The corresponding residual resistivity 
extrapolated from the linear temperature dependence of $\rho_a(T)$  goes from being negative in the nominally pure crystal 
to $\rho_0 = 30~\mu\Omega$~cm in the Zn-doped crystal.
It is clear that the optimally-doped LSCO sample ($x=0.17$) shown in 
Fig.~1, with $\rho_{0}$ = 33 $\mu\Omega$ cm  (as extrapolated from resistivity data above $T_c$),
exhibits much stronger scattering than any of the YBCO samples.  This is true despite the high chemical purity
of the crystal, and is likely a result of the Sr atoms included as dopants acting also as scatterers.
Considering 
all the evidence we
estimate the relative amount of disorder 
in the various crystals studied here to be roughly in the proportion of 100:10:1 for 
LSCO, YSZ-grown YBCO and BZO-grown YBCO, respectively.  

\begin {center}
\begin {table}[t]
\setlength{\extrarowheight}{5pt}
\begin{tabular}{|>{\setlength{\parindent}{1.5mm}}p{2.1cm}|>{\setlength{\parindent}{3mm}}p{1.1cm}|>{\setlength{\parindent}{3mm}}p{1.2cm}|c|c|c|}
\hline
\hspace{0.35cm}Sample&
\hspace{0.1cm}$T_{c}$\hspace{0.3cm}&
\hspace{0.2cm}$p$\hspace{0.25cm}&
$\kappa_{0}/T$\hspace{0.3cm}&
$v_{F}/v_{2}$\hspace{0.2cm}&
$\Delta_{0}$\\
&[K]&&
[$\frac{\mu W}{K^{2}cm}$]&
&
[ meV ]\\
&&&&&\\
\hline
YBCO$_{6.0}$&---&0.0&0$\pm1$&---&---\\
YBCO$_{6.54}$&62&0.10&85$\pm10$&7.9&71\\
YBCO$_{6.6}$&44&0.08&91$\pm13$&8.7&66\\
YBCO$_{6.95}$&93.5&0.16&120$\pm12$&11.5&50\\
YBCO$_{6.99}$&89&0.18&160$\pm12$&15.5&37\\
LSCO 0.05&---&0.05&3$\pm1$&---&---\\
LSCO 0.06 A&5.5&0.06&11$\pm2$&---&---\\
LSCO 0.06 B&8.5&0.06&12$\pm2$&---&---\\
LSCO 0.07&19&0.07&22$\pm2$&1.9&---\\
LSCO 0.09&16&0.09&26$\pm10$&2.4&---\\
LSCO 0.17&34&0.17&$96\pm7$&10.4&---\\
LSCO 0.20&33.5&0.20&$330\pm40$&36&---\\
Bi-2212&89&0.16&$150\pm30$&19&30\\
Tl-2201&15&0.26&$1400\pm70$&270&2\\
\hline
\end{tabular}
\caption{\label{tab}Compilation of $T_c$, doping and residual linear term
in the thermal conductivity as well as values of the quasiparticle
anisotropy ratio
$v_{F}/v_{2}$ from Eq.~2 and gap maximum $\Delta_{0}$ (see caption of
Fig.~6) for the samples in this study.
Data for optimally doped Bi-2212 \cite{Chiao} and overdoped Tl-2201
\cite {Proust} from previous studies are provided for completeness. }
\end {table}
\end {center}
\section{\label{data}Doping Dependence of $\kappa_o/T$}

The low-temperature thermal conductivity of YBCO and LSCO samples is shown as a function of temperature
in Figs. 2 and 3, respectively. The data is plotted as $\kappa/T$ vs $T^2$ because the quantity of interest is
the residual linear term $\kappa_o/T$, defined as the $T=0$ limit of $\kappa/T$, obtained by extrapolation of the 
low-temperature data. This residual linear term can only be due to fermionic carriers and is attributed to zero-energy 
quasiparticles. Indeed, as will be seen below, it is a  direct confirmation, via a robust bulk measurement, 
of the $d$-wave nature of the superconducting order parameter in cuprates. 
The extrapolation procedure is described in detail in the Appendix, 
where the contribution of phonons is analyzed. The main
results of the paper do not depend on the particular extrapolation procedure. 
This is true, for example, for the overall trend with doping, which is immediately evident from Figs. 2 and 3: 
$\kappa_o/T$ decreases steadily with underdoping, 
all the way from the slightly overdoped to the highly underdoped regime.  
Using the extrapolation procedure outlined in the Appendix, the values 
we obtain are given in Table I.
Note that a measurement on a fully deoxygenated YBCO sample with $y=6.0$ correctly yields 
a zero intercept: $\kappa_o/T = 0 \pm 1$ $\mu$W~K$^{-2}$~cm$^{-1}$. 
The values for LSCO agree with those published in a previous study \cite{Ando2}, with the exception of 
their $x = 0.17$ 
sample which has been measured to be approximately twice the value we observe.   We 
attribute this difference to the fact that the crystal studied by Takeya {\it et al.} had a $T_c$ of 
40.2 K compared to our $T_c$ of 34.2 K, pointing to a slightly higher hole concentration (likely
due to different oxygen levels within the crystals).

\begin{figure}
\centering
\vspace{8pt}
\resizebox{\columnwidth}{!}{\includegraphics{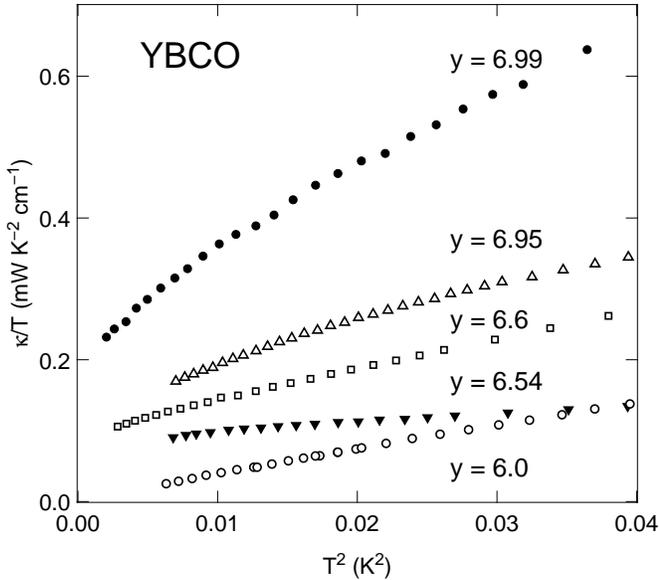}}
\caption{\label{fig:YBCOkappa}Thermal conductivity of $\rm YBa_2Cu_3O_{y}$ plotted
as $\kappa/T$ vs $T^2$.  
Open symbols represent samples grown in YSZ crucibles and 
filled symbols those grown in BZO crucibles.}  
\end{figure}

\begin{figure}
\centering
\vspace{8pt}
\resizebox{\columnwidth}{!}{\includegraphics{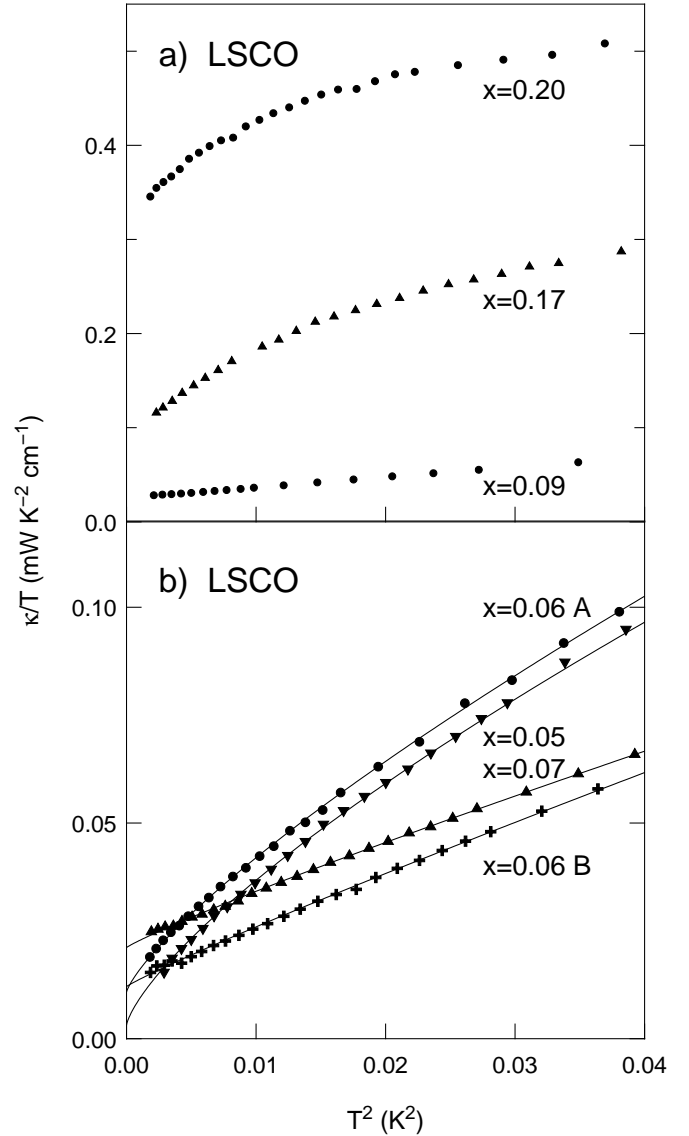}}
\caption{\label{fig:LSCOkappa}Thermal conductivity of La$_{2-x}$Sr$_x$CuO$_4$ plotted
as $\kappa/T$ vs $T^2$, for a) $x=0.09$, 0.17 and 0.20, and 
b) $x=0.05$, 0.06 and 0.07.  The lines through the data are power law fits, discussed
in the Appendix.}
\end{figure}

Let us analyze these results within the framework of standard $d$-wave BCS theory. 
In the clean limit at low temperature, when $k_BT \ll \gamma \ll k_B T_c$, where $\gamma $ is the impurity
bandwidth, the quasiparticle thermal conductivity can be written as: \cite{Durst}

\begin{equation}
 \frac{\kappa_0}{T} = \frac{k_B^2}{3 \hbar} \frac{n}{d} \left( \frac{v_F}{v_2} + \frac{v_2}{v_F} \right),
 \label{eq:koT}
\end{equation}
where $n$ is the number of $\rm{CuO_2}$ planes per unit cell and $d$ is the $c$-axis lattice constant.
$v_F$ and $v_2$ are the quasiparticle velocities normal and tangential to the Fermi surface at the node, respectively,
and are the only two parameters that enter the low-energy spectrum, given by $E~=~\hbar\sqrt{v_F^2k_1^2~+~v_2^2k_2^2}$ 
where ${\bf \hat{k}_1}$ and ${\bf \hat{k}_2}$ are vectors normal and tangential to the Fermi surface at the node, respectively.
The parameter $v_2$ is simply the slope of the gap at the node:

\begin{equation}
  {\bf v_2} = \frac{1}{\hbar}\frac{d\Delta}{d{\bf k}}\Big|_{node} =
              \frac{1}{\hbar k_F}\frac{d\Delta}{d{\phi}}\Big|_{node} = v_2 {\bf \hat{k}_2},
\label{eq:v2}
\end{equation}
where $k_F$ is the Fermi wavevector at the nodal position.
These are  remarkably simple formulae, which provide a direct access to
the parameters that govern low-energy phenomena in a $d$-wave superconductor.
The residual heat conduction in Eq.~2 is not only universal, {\it i.e.} independent of scattering rate (or impurity bandwidth), but it was also
shown to be independent of Fermi-liquid corrections and vertex corrections ({\it i.e.} corrections due to anisotropic 
scattering between nodes). \cite{Durst}  Note, however, 
that the latter two corrections affect the microwave (charge) conductivity (see below).

\begin{figure}
\centering
\resizebox{\columnwidth}{!}{\includegraphics{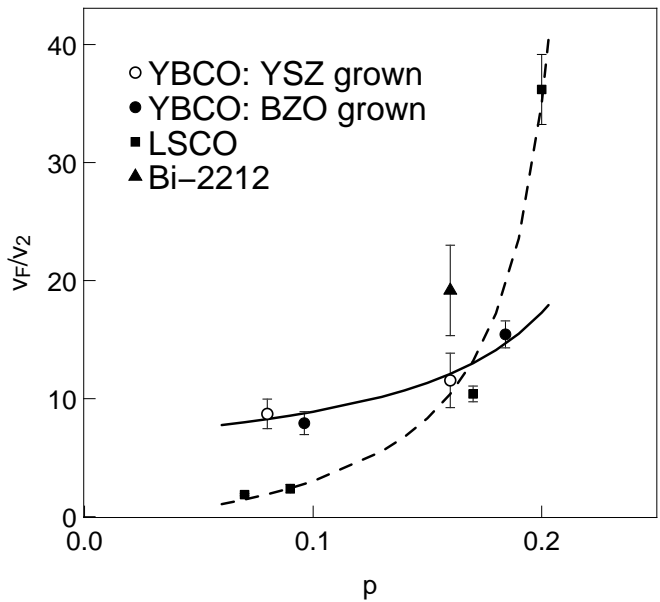}}
\caption{\label{fig:vfov2} Anisotropy ratio $v_F/v_2$, calculated from thermal conductivity data via
Eq.~2, vs hole doping per CuO$_2$ plane, $p$.  The data for Bi-2212 is from Ref.~7.
Lines are guides to the eye (solid for YBCO, dashed for LSCO)}
\end{figure}

 In Fig.~4, the anisotropy ratio $v_F/v_2$ 
is plotted against carrier concentration $p$, using Eq.~2 and the values of $\kappa_0/T$ listed in Table~I. 
Also included is the published value for Bi-2212 at optimal doping. \cite{Chiao}  All three cuprates have a 
comparable anisotropy ratio
at optimal doping: $v_F/v_2$~= 10, 12 and 19, for LSCO, YBCO, and Bi-2212, respectively. It has already been noted 
\cite{Chiao} that
the value of 19 for Bi-2212 is in excellent agreement with the ratio of 20 coming from values of 
$v_F = 2.5 \times 10^7~{\rm cm/s}$ and $v_2 = 1.25 \times 10^6~{\rm cm/s}$ obtained directly from ARPES \cite{Mesot}. 
(Note the value of 12 for the optimally-doped YBCO crystal differs 
slightly -- albeit within error bars -- from our previously published 
result of 14, which was an average of several samples. \cite{Chiao})

\section{\label{sec:discussion}Discussion}

\subsection{Nature of the superconducting order parameter}

Several authors have proposed the existence of a quantum critical
point within the superconducting dome in the phase diagram of 
cuprates, either as a theoretical prediction to explain the diagram
itself or as suggested in various experiments.  Its location is
usually taken to be near (or slightly above) optimal doping, in the
neighbourhood of $p = 0.2$. If it is associated with a change in the
symmetry of the superconducting order parameter, Vojta {\it et al.}
have argued that the most likely scenario is a transition from a pure
$d_{x^2-y^2}$ state to a complex order parameter of the form   
$d_{x^2-y^2} + ix$, where $x$ can have either $s$ or $d_{xy}$ symmetry.
\cite{Sachdev} Sharoni et al. have recently reported a split
zero-bias anomaly in their tunneling on Y-123 thin films as soon as
the material is doped beyond optimal doping, a feature which they
attribute to the appearance of a complex component to the order
parameter in the bulk. \cite{Dagan} The presence of a subdominant
component $ix$ in the order parameter causes the nodes to be removed,
as the gap can no longer go to zero in any direction.  Our observation
of a residual linear term in the thermal conductivity of both YBCO and LSCO,
as well as previous results on optimally-doped Bi-2212 \cite{Chiao} and strongly-overdoped Tl-2201 \cite{Proust},
is a direct consequence of nodes in the gap. It therefore excludes the possibility of any such
subdominant order parameter in the bulk throughout the doping phase diagram.
In other words, if there truly is a quantum critical
point inside the superconducting dome, it does not appear to be associated
with the onset of a complex component in the order parameter.

In view of the ubiquitous nature of the residual linear term in superconducting
cuprates, observed in four different hole-doped materials from strongly-overdoped
Tl-2201 to strongly-underdoped LSCO, two previous results stand out as puzzling anomalies:
the absence of a detectable linear term in electron-doped Pr$_{2-x}$Ce$_x$CuO$_4$ (PCCO)
\cite{Hill} and in hole-doped YBa$_2$Cu$_4$O$_8$. \cite{Hussey}
In particular, note that the upper bound of 0.02 mW~K$^{-2}$~cm$^{-1}$
quoted for $\kappa_{0}/T$ in YBa$_2$Cu$_4$O$_8$ is 4 to 5 times lower than the value obtained here
for YBa$_2$Cu$_3$O$_y$ at a comparable hole concentration ($y=6.54$ or 6.6) - as assessed by the very similar
resistivity curves above $T_c$ - and comparable sample quality. This extremely low value is akin to that found
in non-superconducting strongly-underdoped LSCO ($x=0.05$).

\subsection{Effects of disorder}

One of the most remarkable results of transport theory in $d$-wave superconductors is 
the universal nature of heat conduction, which appears due to a 
cancellation between the increase in normal fluid density and the decrease in mean 
free path observed as the concentration of impurities is increased.\cite{Durst}
This universal behaviour is only found in the clean limit where $\hbar\Gamma \ll \Delta_0$.
In situations where $\Gamma$ is large, (or $\Delta_0$ is 
small), the behavior is no longer universal, and the measured linear term 
may be closer to the normal state value   
$\kappa_N/T$ than the universal limit.   In the extreme case where  $\hbar\Gamma$ 
$\sim$ $\Delta_0$, superconductivity is destroyed and the normal state value of 
$\kappa_N/T$ is recovered.  Therefore the validity of using Eq.~2 
to extract values of $v_F$/$v_2$ from measurements of the 
residual linear term is ensured only when samples are in the clean (universal) limit, $\hbar\Gamma \ll \Delta_0$.  
Universal behaviour in YBCO at optimal doping is already well established, \cite{Taillefer} and inspection of 
Fig. 4  shows that this is confirmed at other dopings.  Indeed, we observe that both BZO and YSZ grown 
crystals yield values of $v_F$/$v_2$ that lie on the same curve despite having an order of 
magnitude difference in purity level, which is strong evidence that the clean limit is reached in our YBCO samples.

In LSCO, the extremely small values of $\kappa_0/T$ measured in highly underdoped samples point to a different 
conclusion.   Indeed, for $x=0.06$, $\kappa_0/T \simeq 12~\mu$W~K$^{-2}$~cm$^{-1}$, while the minimum
value for LSCO allowed by Eq.~2 is  $\frac{k_B^2}{3 \hbar} \frac{n}{d} (1 + 1) = 
18.3~\mu$W~K$^{-2}$~cm$^{-1}$.  The 
data for the LSCO samples with the lowest dopings are plotted in Fig.~5, which shows that the use of 
Eq.~2 for these samples is invalid.   This breakdown suggests that our underdoped LSCO samples are 
not in the clean limit, and hence we cannot extract quantitative information by using Eq.~2, as 
we will do for YBCO in the following sections. The same conclusion would apply to previous LSCO data 
\cite{Ando2}. 

In order to understand the LSCO data within a $d$-wave BCS theory of low 
temperature heat transport, it will be necessary to incorporate the effects of impurity scattering in the 
underdoped regime outside of the clean (universal) limit.  The effect of impurity 
scattering on a $d$-wave superconductor has been worked out in the standard case of a normal state that is 
{\sl metallic}, and conducts heat {\sl better} than the superconducting state. \cite{Maki} When the 
concentration of impurities is increased in 
such a case, $T_c$ is gradually suppressed to zero and the residual linear term 
{\sl rises} monotonically to meet its normal state value.  However, our LSCO samples with $x \leq 0.09$ exhibit 
the well-known insulating upturns in the normal state resistivity associated with the ground state 
metal-insulator transition observed near $x \sim 0.16$. \cite{Boebinger}  
In fact the resistivity in a strong magnetic field appears to 
diverge as $T \to 0$. \cite{Hawthorn}  Thus, for the LSCO samples where $x < 0.16$, 
the effect of increasing the impurity concentration would be to evolve the system towards an {\sl insulating} state, 
or at least one that conducts heat {\sl less well}.  In this scenario, we expect the 
measured residual linear term $\kappa_0/T$ to be {\sl smaller} than the universal value, which 
would explain how in Fig.~5 we measure a linear term smaller than that allowed by Eq.~2.  

 Another possibility is suggested by the theoretical work of Atkinson and Hirschfeld \cite{Atkinson}, in which the 
Bogoliubov-deGennes equations are used to model the paired state as an inhomogenous superfluid. This
approach allows for the possibility of quantum interference processes
such as localization which are neglected in the usual framework.  In their model, the residual linear term $\kappa_0/T$ is 
seen to decrease in the presence of increasing impurity concentration, a direct result of weak localization of carriers.  
The fact that we measure a linear term in underdoped LSCO which is smaller than that allowed by
Eq.~2 may be evidence for the existence of such localization in LSCO.  We hope these observations will stimulate further
theoretical work.

\begin{figure}
\centering
\resizebox{\columnwidth}{!}{\includegraphics{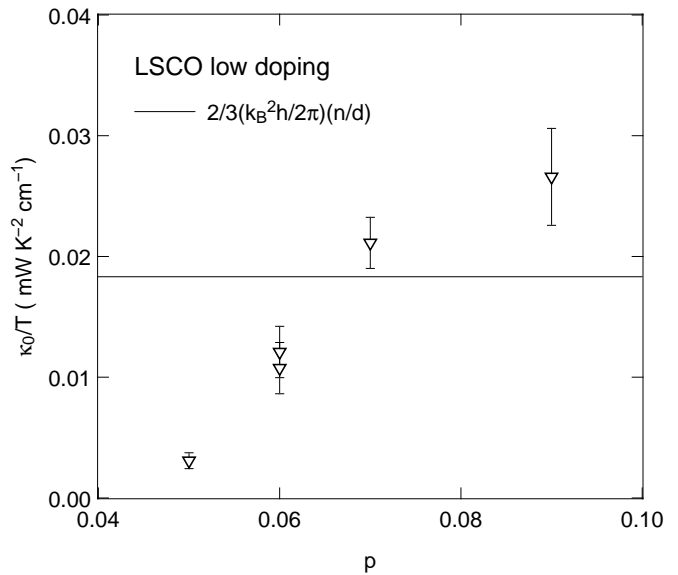}}
\caption{\label{fig:k_0vsp} Measured value of $\kappa_0/T$ for highly underdoped LSCO.  
The solid line represents the minimum possible value allowed by Eq.~2, namely when $v_F/v_2 = 1$.} 
\end{figure}

\begin{figure}
\centering
\resizebox{\columnwidth}{!}{\includegraphics{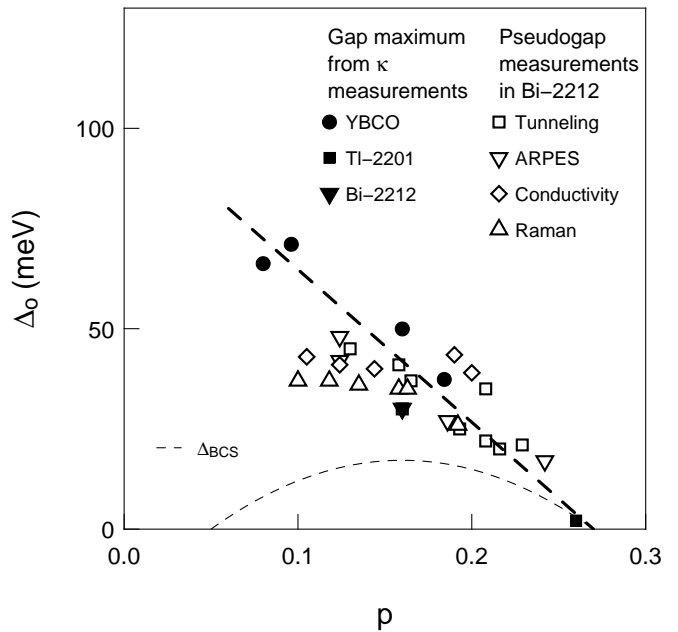}}
\caption{\label{fig:Delta} Doping dependence of the superconducting gap $\Delta_0$ obtained from the 
quasiparticle
velocity $v_2$ defined in Eq.~3 (filled symbols).  Here we assume $\Delta = \Delta_0 {\rm cos}2\phi$, so that
$\Delta_0 = \hbar k_F v_2 / 2$, and we plot data for YBCO alongside Bi-2212 (Ref.~7) and Tl-2201 (Ref.~8).
For comparison, a BCS gap of the form $\Delta_{\rm BCS}=2.14 k_B T_c$ is also plotted,
with $T_c$ taken from Eq.~1 (and $T_c^{max}= 90$~K).
The value of the pseudogap in Bi-2212, as measured by various techniques,
\cite{statt} is also shown (open symbols).  The thick dashed lines is a guide to the eye.}
\end{figure} 

\subsection{Doping dependence of the superconducting gap}

The remarkable success of Eq.~2 at optimal doping validates the extension of our study
across the doping phase diagram, at least for our YBCO samples, where the clean (universal) limit
is established.  In 
interpreting our measurements of the anisotropy ratio $v_F/v_2$ in such a 
study, the first thing to emphasize is the fact that $v_F$, the Fermi velocity at the node, is
essentially independent of doping.
This was shown by ARPES both in Bi-2212 \cite{Mesot} and in LSCO, \cite{shen}
where the slope of the $E$ vs $k$ dispersion
at the Fermi energy is seen to vary by no more than 10\% over the range $0.03 < x < 0.3$, with an average
value of $v_F \simeq 2.5 \times 10^7~{\rm cm/s}$ in both materials. The position of the node in $k$-space is also
independent of doping\cite{Mesot}, with $k_F$ $\simeq$ 0.7~ $\AA~^{-1}$ as measured from 
($\pi$,$\pi$) to the Fermi
surface. As a result, a study of $\kappa_0/T$ vs $p$ yields the doping dependence of $v_2 = v_2(p)$.
In Fig.~6, we plot the slope of the gap at the node as a function of
carrier concentration, not as $v_2$ vs $p$ but in a more familiar guise as the corresponding gap maximum,
$\Delta_0$, of a putative $d$-wave gap function $\Delta$ = $\Delta_0 {\rm cos}2\phi$, via Eqs.~2 and 3.
Given that $k_F$ is constant, this is equivalent to plotting $v_2$ directly.  
The values of $\Delta_0$ are also listed in Table~I.
Again, here we have confined our analysis to YBCO only, given that LSCO was seen to lie outside the clean limit.
Plotted alongside this data is a conventional BCS $d$-wave gap (dashed curve), where 
we have assumed $\Delta_0 = 2.14 k_B T_c$ (weak-coupling approximation).  The $p$ dependence of the gap is 
estimated using Eq.~1, with a maximum $T_c$ at optimal doping of 90 K.

Let us examine the implications of these results by starting on the overdoped side of the phase diagram.
The only available data in the
strongly-overdoped regime is on Tl-2201, \cite{Proust} a single-plane cuprate with optimal $T_c \simeq 90$~K.
For an overdoped crystal with $T_c =15$~K, the measured residual linear term is
$\kappa_0/T = 1.4$~mW~K$^{-2}$~cm$^{-1}$, which yields $v_F/v_2 = 270$ via Eq.~2.
In comparison, the weak-coupling BCS prediction based on the value of
$T_c=15$~K is $v_F/v_2 = 210$, using the values of $v_F$ and $k_F$ given above. 
The good quantitative agreement shows that in this strongly-overdoped regime BCS theory works quite well, and 
the much larger anisotropy ratio is a consequence of the much smaller $T_c$.

We now turn our attention to the underdoped region of the phase diagram. In the case of 
YBCO the decrease in $\kappa_0/T$ by a factor 2 between $y=6.99$ and $y=6.54$ provides one of the main results of this paper: 
the velocity ratio decreases with underdoping; it drops from
16 to 8 in going from a sample with $T_c=89$~K to an underdoped sample with $T_c=62$~K. {\sl This reflects an underlying
steepening of the gap at the node while $T_c$ drops, with underdoping}. (Note that this is in contradiction with 
the results of Mesot {\it et al.} who extracted a slope of the gap from their ARPES measurements on Bi-2212 near optimal
doping that seemed to decrease slightly with underdoping, \cite{Mesot} and with the analysis of Panagopoulos {\it et 
al.} who extract a gap maximum from their penetration depth measurements that remains approximately constant in the underdoped 
regime.\cite{pana}) 

Taken by itself, this could be attributed either to a gradual departure from weak-coupling towards strong-coupling
BCS superconductivity, with a growing ratio $\Delta_0/T_c$. It could also be interpreted as a gradual deformation
of the gap shape, from a simple cos$2\phi$ angular dependence to a much steeper function with a decreasing average
gap that scales with $T_c$.
However, in view of the known behaviour of the pseudogap,  
these explanations are unlikely to be the main story.
Indeed, the growth of the low-energy gap observed through $\kappa_0/T$ is highly reminiscent 
of the similar trend observed in the high energy gap ( or pseudogap ) with underdoping.
In fact the growth of $\Delta_0$ derived from $v_2$ is in quantitative agreement with the 
pseudogap maximum determined by ARPES, \cite{norman,white,loeser} tunneling 
\cite{renner,dewilde} $a-b$ plane optical conductivity \cite{puchkov} and Raman scattering, \cite{blumberg,hackl} 
as shown in Fig.~6.


   This striking similarity in scaling points to a common origin, which allows us to say 
the  following things on the nature of the pseudogap.  First, due to the 
very existence of a residual linear term, the (total) gap seen in thermal 
conductivity at $T \to 0$ is one that 
must have nodes.
Secondly, it has a linear dispersion as in a $d$-wave gap ({\it i.e.} it has a Dirac-like spectrum). 
Thirdly, it is a quasiparticle gap and not just a spin gap. 
A fundamental question is whether the pseudogap is related to or independent of superconductivity.
The first and most natural possibility is that it is due to some form of precursor 
pairing. A second possibility is that it may come from a distinct non-superconducting state.
Indeed, a universal thermal conductivity is also possible in a non-superconducting state as long as the 
energy spectrum is Dirac-like ({\it i.e.} linear dispersion).  For example, a universal (charge) conductivity
was derived for a degenerate semiconductor in 2D. \cite{fradkin}
Interestingly, the $d$-density-wave (DDW) state proposed as an explanation for the pseudogap phenomena seen in 
underdoped cuprates \cite{chakravarty} also exhibits a universal conductivity provided that the chemical 
potential $\mu$ = 0. In the 
region where both orders coexist -- DDW
and $d$-wave superconductivity (DSC) -- Eq.~2 is then predicted to hold, \cite{Nayak} with $v_2$ replaced by 
$\sqrt{(v_{\Delta}^{DDW})^2~+~(v_{\Delta}^{SC})^2}$, 
where $v_{\Delta}^{DDW}$ and $v_{\Delta}^{SC}$ are the gap velocities for the two types of order, respectively.
The main question then is how does the chemical potential evolove as a function of doping.

    In summary, our measurements of $\kappa/T$ throughout the 
phase diagram allow us to make the following statements about the evolution of $\Delta_0$ with doping.  First, the 
extrapolated value of the gap maximum from thermal conductivity in the overdoped regime is
in excellent quantitative agreement with that expected from BCS theory.  
Secondly, $\Delta_0$ continues to grow with underdoping while 
$T_c$ rises and then falls, in contradiction to what one would expect from BCS theory.  
The divergence of these two energy scales in the underdoped regime 
is a manifestation of the pseudogap, whose presence is now revealed at very low energies
in a bulk measurement on crystals of the utmost quality and purity.  
The fact that the gap preserves its pure $d$-wave form (with nodes on the Fermi surface) 
throughout strongly suggests that the pseudogap is superconducting in origin.

\subsection{Superfluid density and microwave conductivity}

One way to shed further light on the nature of the low-energy electron state in underdoped YBCO is to compare heat 
transport and charge 
dynamics. For a $d$-wave BCS superconductor, Durst and Lee have shown that the two conductivities are affected
differently by scattering anisotropy and quasiparticle interactions. \cite{Durst} 
The charge conductivity in the $\omega \to 0$ and $T \to 0$ limit is given by

\begin{equation}
{\rm lim_{~T \to 0}} \hspace{3pt}\sigma_1 ( T ) = \sigma_0 = 
\frac{e^2}{\hbar} \frac{1}{\pi^2} \frac{n}{d}~\beta_{VC} ~\alpha_{FL}^2 ~\frac{v_F}{v_2},
 \label{eq:sigma_1}
\end{equation}
where $e$ is the electron charge. The factor $\beta_{VC}$ is due to vertex corrections and is greater than 1.0 when
impurity scattering is anisotropic. This simply reflects the fact that intra-node scattering will degrade a charge current
less than inter-node (opposite- or side-node) scattering that involves a larger change in momentum. 
This is the discrete version of the $(1 - {\rm cos}\theta)$ term that enters normal state conductivity and reflects 
the predominance of back-scattering over small-angle scattering. Numerical calculations suggest that $\beta_{VC}$ can be 
large ({\it e.g.} in excess of 10) in high-purity samples as long as impurity scattering is in the unitary 
limit. \cite{Durst}  Note that vertex corrections have negligible effect on heat transport.
The factor $\alpha_{FL}^2$ is a Fermi-liquid correction
which arises because of quasiparticle-quasiparticle interactions.  
The same factor also enters in the low-temperature slope of the normal fluid density $\rho_n(T)=\rho_s(T=0)-\rho_s(T)$: 
\cite{Durst}

\begin{equation}
 \frac{\rho_n(T)}{m}  = \frac{2\rm\hspace{2pt}{ln}2}{\pi} \frac{1}{\hbar^2} \frac{n}{d} ~ \alpha_{FL}^2 ~\frac{v_F}{v_2} ~k_B 
T .
 \label{eq:superfluid}
\end{equation}
The temperature dependence of the $a$-axis superfluid density  of YBCO crystals very similar to ours was measured
via the penetration depth. \cite{Hosseini,Turner} The value of $\alpha_{FL}^2 \frac{v_F}{v_2}$ obtained from 
this data is shown in Fig.~7. Using the value of $v_F/v_2$ from $\kappa_0/T$ (averaging the YSZ-grown and 
BZO-grown data) yields

\begin{figure}
\centering
\resizebox{3.4in}{!}{\includegraphics{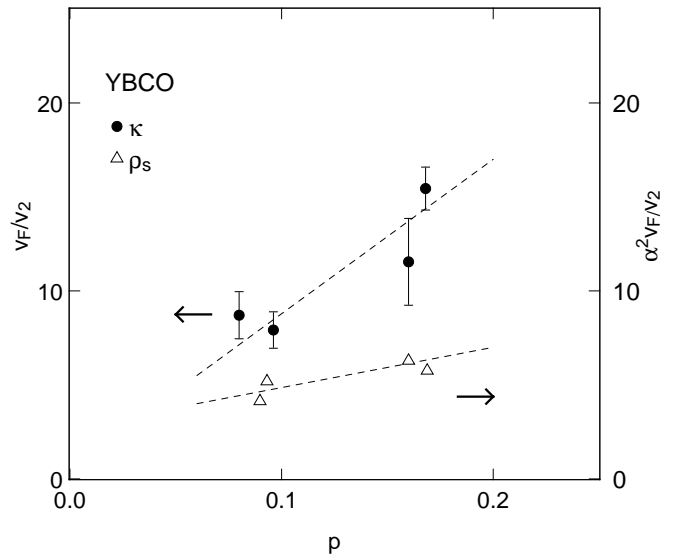}}
\caption{\label{fig:vfov2} Quasiparticle velocity ratio in YBCO obtained from universal heat transport, as
$v_F/v_2$ (circles) via Eq.~2, and from superfluid density data of Refs.~16 and 42, as
$\alpha_{FL}^2 \frac{v_F}{v_2}$ (triangles) via Eq.~5.
Lines are guides to the eye.}  \end{figure}

\begin{equation}
 \alpha_{FL}^2 \simeq 0.4 - 0.5~, ~~~~~ {\rm  at}~p \simeq 0.16
 \label{eq:alpha-1}
\end{equation}

\begin{equation}
 \alpha_{FL}^2 \simeq 0.6 - 0.7~, ~~~~~ {\rm  at}~p \simeq 0.09
 \label{eq:alpha-2}
\end{equation}

A similar value was previously derived for optimally-doped Bi-2212. \cite{Chiao} 
We conclude that this FL parameter is near unity and, more importantly, is only weakly dependent on doping.
 In a recent paper, Ioffe and Millis \cite{ioffe_millis} argue that a 
doping independent $\alpha_{FL}^2$, interpreted as effective charge, is inconsistent with both the Brinkman-Rice 
mean field 
theory and slave boson gauge theory approaches to the Mott physics of high-$T_{c}$ 
materials.  Indeed the combination of a doping independent $\alpha_{FL}^2$ and $v_{F}$, 
along with a $v_{2}$ that increases with decreasing doping provides a significant challenge to 
microscopic theories of $d$-wave superconductivity in cuprates.

The microwave conductivity $\sigma_1(\omega,T)$ of YBCO was recently
measured in crystals nominally identical to ours
with $y=6.50$ \cite{Turner} and $y=6.99$. \cite{Hosseini}
Even though the measurements go down to 1 GHz and 1.3 K, it turns out to
be unclear how to reliably extrapolate this data
to the $\omega = 0$ and $T=0$ limit, so that a meaningful comparison of
$\kappa_0/T$ and $\sigma_0$ is not quite possible at this stage.
The shape and temperature dependence of the microwave spectrum for the
$y=6.50$ sample for example is suggestive of non-unitary scattering
close to the Born limit, implying that the low-temperature universal limit
regime may not be reached by 1.3 K.  Further work is needed to ascertain
whether this is indeed the correct scenario.


\section{Conclusions}

We have studied the evolution of thermal transport as $T \rightarrow 0$ in the cuprate superconductors YBCO and 
LSCO over a wide range of the doping phase diagram. The residual linear term, $\frac{\kappa_{0}}{T}$, 
is observed to be finite throughout the superconducting region. This proves that the gap always
has nodes on the Fermi surface, a fact that has two implications:
1) it rules out the possibility of a multicomponent order parameter of the type $d+ix$ in the bulk, appearing at a putative 
quantum phase transition, and
2) it argues strongly in favour of a superconducting origin to the pseudogap ({\it e.g.} precursor pairing).
As the Mott insulator is approached, $\frac{\kappa_{0}}{T}$ is observed to 
decrease, leading to a decreasing value of the quasiparticle velocity anisotropy ratio, $v_F/v_2$.  This result 
offers some of the first insights into the doping dependence of several important 
quasiparticle parameters.  
First, the slope of the $d$-wave superconducting gap at the nodes, $v_2$, is seen to 
increase steadily as doping is decreased, consistent with a growth of the gap in the underdoped regime.  This is 
in contradiction to what one naively expects from BCS theory, where the gap scales with $T_c$.  
The gap we extract at very low energies follows closely the pseudogap measured mostly at much higher energies 
by other techniques. This close tracking of the pseudogap shows that the gap remains roughly of the pure $d$-wave form 
throughout the phase diagram.
Secondly, a comparison with superfluid density reveals that the 
quasiparticle effective charge is weakly dependent on doping and close to unity.

The considerable difference between the magnitude of the change in  $\frac{\kappa_{0}}{T}$ with underdoping 
for the LSCO and YBCO samples provides clues as to the role of disorder in the underdoped 
regime.  In particular, the small value of the residual linear term of the most highly underdoped LSCO samples is 
incompatible with the standard theory of transport for $d$-wave superconductors, motivating theoretical work which 
would incorporate the effects of impurities in a superconductor whose normal state is insulating rather than metallic in 
nature.

\section{Acknowledgements}

We acknowledge stimulating discussions with A. Millis and M. Franz, and thank H.
Dabkowska, G. Luke, K. Hirota and B. Gaulin for assistance in the growth 
of LSCO. We also thank W. A. MacFarlane and P.  Fournier for assistance in 
various aspects of the measurements.  This work was supported by the 
Canadian Institute for
Advanced Research and funded by NSERC of Canada.  C.L. acknowledges the
support of an FCAR scholarship and D.H. thanks the Walter Sumner 
Foundation.  \\

\bibliography{dopingpaper}

\vspace{0.5cm}
$^{\star}$ Present address: Laboratoire National des Champs 
Magn\'etiques Puls\'es, 143 avenue de Rangueil, 31432 Toulouse, France.\\
\\
$^{\dagger}$ Present address: Department of Physics, University of California 
at Berkeley, Berkeley, California 94720

\include{dopingpaper.bib}

\section{Appendix: Phonon Thermal Conductivity in d-wave Superconductors}

In order to use thermal conductivity as a direct probe of low-energy quasiparticles in $d$-wave 
superconductors,
the contribution from phonons must be reliably extracted.  This may be achieved by performing experiments in 
the
regime
${~T \to 0}$, where the phonon mean free path becomes limited only by the physical dimensions of the sample.
From
simple kinetic theory, the conductivity of phonons in this boundary-limited scattering regime is given by

\begin{equation}
 \kappa_{ph} = \frac{1}{3} \beta <\hspace{-3pt}v_{ph}\hspace{-3pt}> \Lambda_0 T^3 \label{eq:kappa_ph}
\end{equation}

where $\beta$ is the coefficient of phonon specific heat, $\Lambda_0$ is the temperature-independent mean free 
path, and $<\hspace{-3pt}v_{ph}\hspace{-3pt}>$ is a suitable average of the acoustic sound velocities.   The
electronic linear term is then naturally
extracted by plotting thermal conductivity data as $\frac{\kappa}{T}$ vs $T^2$ and interpreting the intercept 
as the
residual linear term at $T = 0$, and the slope as the phonon contribution governed by Eq.~8.
The extension of our
measurements into the highly underdoped
region of the cuprate phase diagram, where $\frac{\kappa}{T}$ becomes very small,
led us to refine this extrapolation
technique.

\begin{figure}[t]
\centering
\resizebox{3.3in}{!}{\includegraphics{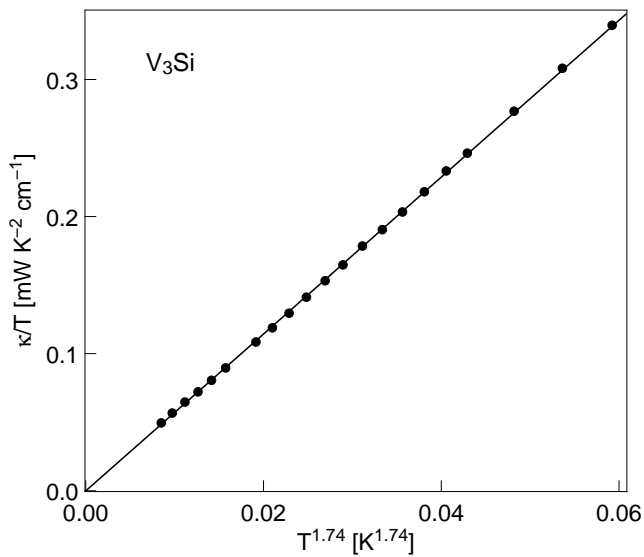}}
\caption{\label{fig:v3sipowerlaw} Thermal conductivity of the $s$-wave
superconductor V$_3$Si.  The data is plotted as
$\frac{\kappa}{T}$ vs $T^{1.74}$, and the line represents a free fit to
the data of the form of Eq.~9.  The resulting linear term is
zero: 0
$\pm$ 1 $\mu$W~K$^{-2}$~cm$^{-1}$, consistent with that
expected for a nodeless superconductor.} \end{figure}

\begin{figure}[b]
\centering
\resizebox{3.3in}{!}{\includegraphics{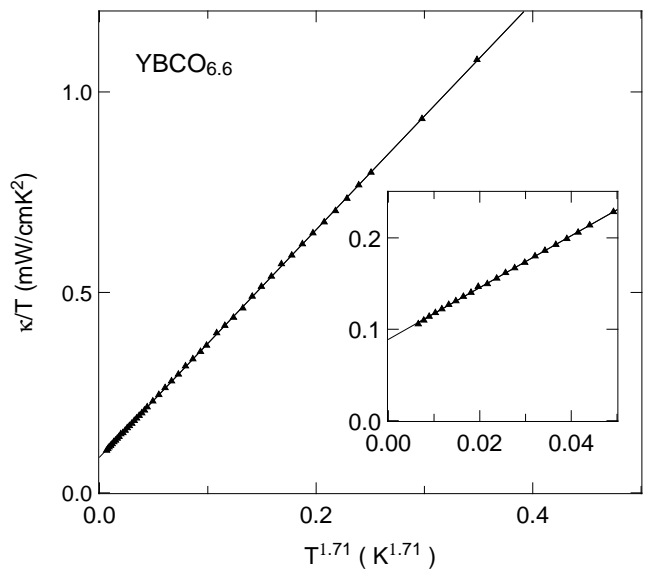}}
\caption{\label{fig:undOpowerlaw} Thermal conductivity of YBCO$_{6.6}$.
The data is plotted as $\frac{\kappa}{T}$ vs $T^{1.71}$, and the line is a linear fit.
{\sl Inset}: zoom at low temperatures. Note the presence of a clear residual linear term, the contribution of nodal quasiparticles.}
\end{figure}

To motivate why this may be necessary, consider the possible scattering
mechanisms available to a phonon impinging upon the surface of a crystal.  The phonon may either be absorbed 
and
reemitted with an energy distribution given by the local temperature (diffuse scattering) or it may be 
reflected elastically (specular reflection).  In the case of diffuse scattering, the phonon is reradiated in a random 
direction
resulting in a temperature independent value of $\Lambda_{0}$ and a $T^3$ dependence of $\kappa_{ph}$ as recognised by
Casimir. \cite{Casimir}   However, as the temperature of a crystal is reduced and the average phonon
wavelength increases,
a surface of a given roughness appears smoother, which may increase the occurance of specular reflection and result in
a mean free path which varies as some power of temperature, so that $\kappa_{ph}$ $\propto$ $T^{\alpha}$.
We would thus
expect a
deviation from the diffuse scattering limit of $T^3$ temperature dependence of $\kappa_{ph}$ for samples with   
sufficiently smooth surfaces.  Such an effect has been previously observed in many studies of low-temperature
phonon heat transport in high-quality
crystals, such as Al$_2$O$_3$, \cite{Pohl} Si, \cite{Hurst}  KCl and KBr, \cite{Seward} LiF \cite{Thatcher}
and diamond. \cite{BSZ}

This effect can be seen in most of our samples, manifesting itself as a gradual curvature in the low temperature
part of our data when plotted as $\frac{\kappa}{T}$ vs $T^2$ (see Figs.~2 and 3).  
In light of this, the thermal conductivity in the boundary scattering regime
is more correctly modelled as:

\begin{equation}
 \kappa = \kappa_{el} + \kappa_{ph} = A T + B T^\alpha\label{eq:kappa_pow}
\end{equation}
with $\alpha < 3$.
Here $A$ is the coefficient of the electronic linear term, and $B$ the temperature-independent coefficient of the phonon
term, where $\alpha$ is some power of temperature, typically between 2 and 3.
(Note that there is no fundamental reason for a single power law - it is simply an empirical result.
For example, in Al$_2$O$_3$ previous studies \cite{Pohl} have found $\alpha=2.77$.)

In superconductors possessing an isotropic or $s$-wave gap, the absence
of an electronic linear term at low temperatures reveals this effect well.  Plotted in Fig.~8 is
thermal conductivity data for the $s$-wave superconductor V$_3$Si, \cite{boaknin} where the 
line is the result of a free fit to
a simple power law as in Eq.~9.  Such a procedure yields a linear term $A$ = -0.04 $\pm$ 1
$\mu$W~K$^{-2}$~cm$^{-1}$, a phonon coefficient $B$
= 5.73 $\pm$ 0.07 mW~K$^{-(\alpha +1)}$cm$^{-1}$, and an exponent $\alpha$ = 2.74 $\pm$ 0.01.  The
validity of such a  fitting procedure is best seen
by plotting the data as in Fig. ~8, with the $x$-axis in units of $T^{\alpha - 1}$.  The
striking linearity of the data on this plot, and the fact that it extrapolates to zero, is good evidence for the appropriateness of 
the power law fitting procedure.

Fig.~9 shows the results of a fit to Eq.~9 for our YBCO$_{6.6}$ crystal, where a power law of $\alpha$ = 2.71 is seen
to persist to temperatures as high as 550 mK.
Power-law fits are also shown in Fig.~3b, this time on a $\kappa/T$ vs $T^2$ plot, for underdoped LSCO samples.
The value of $\alpha$ observed in our samples was found to vary over a range from 2.4 for the YBCO $y=6.99$ crystal to
2.92 for the LSCO $x=0.09$ crystal.

It is worth stressing that the single-power-law fitting procedure described here is simply an
empirical approach to extrapolate the most reliable value of $\kappa/T$ at $T=0$. As a three-parameter
{\it free} fit to the data over a temperature range typically of a decade (50 - 500 mK), it is far
better than the old two-parameter {\it forced} fit to a $\kappa/T = a + bT^2$ form, which invariably   
must be limited to the very lowest temperatures (usually below 150 mK or so)
and typically overestimates the value of $\kappa_0/T$.
However, it must be noted that in some cases it doesn't work well over the whole range up to $\sim$500 mK.
This is indeed the case in our LSCO samples $x=0.17$ and $x=0.20$, where the single-power law fit is
inadequate to describe the rapid fall of  $\kappa/T$ below 150 mK. (Such a decrease was also observed
in LSCO samples with similar doping levels in a previous study \cite{Ando2}.) The low temperature drop is
most likely in the electronic channel ($\kappa_e(T)$), but its origin is as yet unclear. A downturn in
$\kappa/T$ at temperatures below 0.2 K or so can be induced in a sample on purpose by simply using
highly resistive contacts (k$\Omega$ or higher). The drop is then attributed to the rapid deterioration
of the coupling between electrons and phonons at those very low temperatures. In such a case, the
extrapolation procedure must be based only on data above the downturn.
It is not clear that the same phenomenon can still occur in the presence of excellent contacts,
like those used here (less than 1 $\Omega$).
These considerations are explored and discussed more fully elsewhere \cite{future_work}.
In conclusion, when a single-power law works over a wide range of temperature ({\it e.g.} up to 0.5 K),
then the extrapolation is reliable; if it doesn't work, then one needs to understand why and may be forced
to rely only on data above any anomalous downturns.  

In this paper, all data was successfully analyzed using the power-law procedure,
except for LSCO samples $x=0.17$ and $x=0.20$, where instead a linear fit to the data in Fig.3a
was used below 150 mK, yielding the values of $\kappa_0/T$ quoted in Table I. Use of a power-law
fit above 150 mK yields higher values, namely $\kappa_0/T = 0.2$ and $0.4$~mW~K$^{-2}$~cm$^{-1}$ respectively,
which has no impact on any of our conclusions.

\end{document}